\documentclass[aps, prl, twocolumn, amsmath, amssymb, nofootinbib, superscriptaddress]{revtex4-2}

\usepackage{amsmath}
\usepackage{amsfonts}
\usepackage{amssymb}
\usepackage{mathrsfs}
\usepackage{bm}
\usepackage{braket}
\usepackage{xcolor}
\usepackage{tikz}
\usetikzlibrary{calc}
\usetikzlibrary{positioning}
\usepackage{graphicx}
\usepackage{float}
\usepackage[caption=false]{subfig}
\usepackage{overpic}
\usepackage{dsfont}
\usepackage{hyperref}
\hypersetup{hidelinks}
\usepackage[capitalise, nameinlink]{cleveref}
\usepackage{soul}
\usepackage{mathtools}

\begin{document}
    \title{Classical symmetry enriched topological orders and distinct monopole charges \\ 
    for dipole-octupole spin ices}
    
    \author{Pengwei Zhao}
    \affiliation{International Center for Quantum Materials, Peking University, Beijing 100871, China}

    \author{Gang v.~Chen} 
    \email{chenxray@pku.edu.cn}
    \affiliation{International Center for Quantum Materials, Peking University, Beijing 100871, China}
    \affiliation{Tsung-Dao Lee Institute, Shanghai Jiao Tong University, Shanghai 201210, China}
    \affiliation{Collaborative Innovation Center of Quantum Matter, Beijing 100871, China}
    \affiliation{The University of Hong Kong, Shenzhen Institute of Research and Innovation, Shenzhen 518057, China}
    \affiliation{Beijing Key Laboratory of Quantum Devices, Peking University, Beijing 100871, China}
    \date{\today}

\begin{abstract}
Distinct symmetry enriched topological orders often do not have classical distinctions. Motivated by the recent progress on the pyrochlore spin ice materials based on the dipole-octupole doublets, we argue that the dipolar spin liquid and the octupolar spin liquid can be distinguished through the magnetic charges of the magnetic monopoles in the classical spin ice regime. It is observed and predicted that the long-range dipole-dipole interaction renders the magnetic monopole of the dipolar spin ice a finite magnetic charge via the dumbbell picture even in the classical regime. For the octupolar spin ice, however, a zero magnetic charge is expected from this mechanism in the classical regime. We expect this smoking-gun observation to resolve the debate on the nature of Ce$_2$Sn$_2$O$_7$, and more broadly, this work may inspire further experiments and thoughts on the Ce-pyrochlore spin liquids, Nd-pyrochlore antiferromagnets, Er-based spinels, and the distinct properties of the emergent quasiparticles in various symmetry enriched topological phases.
\end{abstract}

\maketitle

\noindent\emph{Introduction.}---Intrinsic topological orders are described by topological quantum field theories with deconfined fractionalized excitations and fractional statistics \cite{wen_colloquium_2017,sachdev_topological_2018,nayak_non-abelian_2008,wenQuantumOrdersSymmetric2002}. Topologically ordered phases are extremely scarce in nature. The only known examples are the fractional quantum Hall effect \cite{stormer_fractional_1999,tsui_two-dimensional_1982}, and more recently, the fractional quantum anomalous Hall effect~\cite{cai_signatures_2023,xu_observation_2023}. Due to the charge conservation symmetry, anyonic excitation in the quantum Hall liquids can carry a fractional charge that has been confirmed in the shot noise experiments~\cite{de-picciottoDirectObservationFractional1997}. Thus, extra symmetries render nontrivial quantum numbers to the emergent quasiparticle excitations in the topologically ordered states and thereby make the topological orders more experimentally visible. Fundamentally, symmetry could enrich the topological order and generate distinct quantum phases with the same topological order~\cite{wenQuantumOrdersSymmetric2002,mesarosClassificationSymmetryEnriched2013,PhysRevB.87.104406}. There have been quite a few systematic classifications of the interplay between the symmetry and topological order in recent years~\cite{mesarosClassificationSymmetryEnriched2013,luClassificationPropertiesSymmetryenriched2016}. Despite these theoretical efforts, the physical examples that realize these symmetry enriched topological orders are extremely rare.

The fractional quantum Hall states and the related fractional Chern insulators are arguably the only realistic and accepted examples of intrinsic topological orders. The $\mathrm{U}(1)$ charge conservation renders the anyonic particles with fractional charges. The charge-carrying nature of the anyons allows for the detection of the anyons with the charge-sensitive probes. With a somewhat similar but not exactly the same spirit, the emergent ``magnetic monopole'' acquires an effective magnetic charge in the dipolar spin ice from the long-range magnetic dipole-dipole interaction~\cite{castelnovoMagneticMonopolesSpin2008}. While the spin ice physics could simply emerge with the nearest-neighbor Ising interaction on the pyrochlore lattice, the emergent magnetic charge of the monopole is a natural gift from the long-range dipole-dipole interaction \cite{castelnovoMagneticMonopolesSpin2008,denhertogDipolarInteractionsOrigin2000,isakovWhySpinIce2005}. As the magnetic dipole moments, the spins naturally have the magnetic dipole-dipole interaction in addition to the exchange interaction. This extra dipole-dipole interaction does not suppress the {\sl finite-temperature} spin ice physics, and hence, the spin ice with the dipole-dipole interaction is sometimes referred to as dipolar spin ice. More importantly, it introduces the effective magnetic charge to the ``magnetic monopole'' and thus makes the ``magnetic monopole'' visible in the magnetic-charge sensitive probes \cite{castelnovoMagneticMonopolesSpin2008,isakovWhySpinIce2005,ladakDirectObservationMagnetic2010,dusadMagneticMonopoleNoise2019,kadowakiObservationMagneticMonopoles2009}. If the spins are not magnetic dipole moments, there will not be a magnetic dipole-dipole interaction, and the magnetic monopole will not acquire any effective magnetic charge. This is an interesting piece of classical physics and resembles the charge fractionalization in the fractional quantum Hall effect.

Given the above background, we raise and answer the following question in this paper. Can distinct symmetry enriched topological phases be distinguished in the classical limit or classically? We do not have a general answer to this question. Instead, we address it with a specific case for the three-dimensional (3D) $\mathrm{U}(1)$ topological orders in the context of pyrochlore spin ice~\cite{PhysRevB.69.064404}. It was previously proposed that the dipole-octupole doublets on the pyrochlore lattice could realize two distinct symmetry enriched $\mathrm{U}(1)$ topological orders~\cite{huangQuantumSpinIces2014,liSymmetryEnrichedU12017}, i.e., the dipolar $\mathrm{U}(1)$ spin liquid and the octupolar $\mathrm{U}(1)$ spin liquid~\cite{smithExperimentalInsightsQuantum2025}. In addition to the parameter regimes for their appearance~\cite{huangQuantumSpinIces2014}, their physical properties are discussed to distinguish the dipolar and octupolar $\mathrm{U}(1)$ spin liquids~\cite{huangQuantumSpinIces2014,liSymmetryEnrichedU12017,Bhardwaj_2022}. In addition to the distinct spin correlations in these different symmetry enriched spin liquids~\cite{huangQuantumSpinIces2014}, the selective measurement of the spinon continuum for the octupolar $\mathrm{U}(1)$ spin liquid plays an important role in understanding the spectroscopic measurements~\cite{liSymmetryEnrichedU12017,PhysRevResearch.2.013334,PhysRevResearch.5.033169,zhaoInelasticNeutronScattering2024}. The relevance of the dipole-octupole doublet to the Nd-based pyrochlores and others was pointed out much earlier~\cite{huangQuantumSpinIces2014,smithExperimentalInsightsQuantum2025,PhysRevB.94.104430}, but these materials are known to be magnetically ordered despite the interesting moment fragmentation~\cite{rauFrustratedQuantumRareEarth2019,petitObservationMagneticFragmentation2016,xuOrderOutCoulomb2020,xuAnisotropicExchangeHamiltonian2019,PhysRevB.94.104430}. The connection of the dipole-octupole doublet to the Ce pyrochlore spin liquid material~\cite{sibilleCandidateQuantumSpin2015,smithExperimentalInsightsQuantum2025} was clarified a bit later by one of the authors and collaborator~\cite{liSymmetryEnrichedU12017}, and received some further theoretical attention including our own efforts~\cite{PhysRevResearch.2.013334,PhysRevResearch.5.033169,PhysRevResearch.2.013066,zhaoInelasticNeutronScattering2024,PhysRevResearch.2.023253,PhysRevLett.132.066502,Bhardwaj_2022,PhysRevLett.129.097202}. After this progress, the next-level question is to understand which spin liquid in the phase diagram of the dipole-octupole doublet is realized in the Ce pyrochlores. In particular, there is an ongoing debate on the nature of the ground state for Ce$_2$Sn$_2$O$_7$, and the ground states for the other Ce pyrochlores such as Ce$_2$Zr$_2$O$_7$ and Ce$_2$Hf$_2$O$_7$ remain to be understood~\cite{PhysRevLett.122.187201,Gao_2019,smithCaseQuantum2022,Gao_2022,PhysRevX.15.021033,PhysRevB.108.054438,PhysRevB.108.174411,smithExperimentalInsightsQuantum2025,poree2024dipolaroctupolarcorrelationshierarchyexchange,smithTwoPeakHeatCapacity2025,pottsExploitingPolarizationDependence2024}. References~\onlinecite{Sibille_2020,Por_e_2024} suggested an octupolar $\mathrm{U}(1)$ spin liquid for Ce$_2$Sn$_2$O$_7$, and Reference~\onlinecite{yahneDipolarSpinIce2024} worked on a different sample and proposed Ce$_2$Sn$_2$O$_7$ to have an ordered ground state at experimentally inaccessible temperatures but is located in the dipolar spin ice regime. Our proposal for resolving these debates is that the dipolar (octupolar) spin ice has a finite (zero) magnetic charge for the magnetic monopole in the classical spin ice regime.

\noindent\emph{Model}.---We start with the dipole-octupole doublets of the Ce$^{3+}$ ions in the Ce-pyrochlore materials~\cite{smithExperimentalInsightsQuantum2025,
liSymmetryEnrichedU12017,huangQuantumSpinIces2014}. Ground states of the Ce$^{3+}$ ion here are identified as a dipole-octupole doublet, described by an effective spin-$1/2$ operator $\bm{\tau}$~\cite{liSymmetryEnrichedU12017}. Under the space group symmetry, $\tau^x$ and $\tau^z$ transform as magnetic dipole moments while $\tau^y$ transforms as a magnetic octupole moment. The magnetic moment is ${\bm{\mu}_i=g\mu_{\text{B}}\tau^z_i\hat{\bm{z}}}$ of Ce$^{3+}$, where $g$ is the Land\'e $g$ factor.
The Hamiltonian of the Ce-based pyrochlore spin ice is given by~\cite{huangQuantumSpinIces2014,liSymmetryEnrichedU12017}
\begin{equation}
\label{eq1}
	\begin{aligned}
		H&=\sum_{\braket{ij}}[J_x\tau^x_i\tau^x_j+J_y\tau^y_i\tau^y_j+J_z\tau^z_i\tau^z_j\\
		&+J_{xz}\left(\tau^x_i\tau^z_j+\tau^z_i\tau^x_j\right)]-\sum_i(\hat{\bm{z}}_i\cdot\hat{\bm{e}})hg\mu_{\text{B}}\tau^z_i\\
		&+\frac{1}{2}\sum_{i,j}^{r_{ij}>r_{\text{nn}}}\frac{\mu_0 g^2\mu_{\text{B}}^2}{4\pi}\frac{\hat{\bm{z}}_i\cdot\hat{\bm{z}}_j-3(\hat{\bm{z}}_i\cdot\hat{\bm{r}}_{ij})(\hat{\bm{z}}_j\cdot\hat{\bm{r}}_{ij})}{r_{ij}^3}\tau^z_i\tau^z_j.
	\end{aligned}
\end{equation}
The first two rows include all possible symmetry-allowed couplings for the nearest neighbors and the external magnetic field $h\hat{\bm{e}}$. The last row is the dipole-dipole interaction between these magnetic dipole moments, $\mu_0$ is the permeability of the vacuum, and $r_{\text{nn}}$ is the distance of nearest neighbors (NN). In our choice, the dipole-dipole interaction starts from the second nearest neighbor to infinity. Thus, the first line has already included the NN contribution from the dipole-dipole interaction.

\begin{table}[t]
	\centering
	\begin{tabular}{cccc}
		\hline\hline
		Easy axis & spin ice & $\tilde{g}$ & Magnetic charge $Q_{\text{m}}=2q_{\text{m}}$ \\
		\hline
		$x$ & dipolar & $-g\sin\theta$ & $-\sqrt{2/3}g\mu_{\text{B}}\sin\theta /r_{\text{nn}}$  \\
		$y$ & octupolar & $0$ & 0 \\
		$z$ & dipolar & $g\cos\theta$ & $\sqrt{2/3}g\mu_{\text{B}}\cos\theta /r_{\text{nn}}$ \\
		\hline\hline
	\end{tabular}
	\caption{The effective $g$ factor $\tilde{g}$ and magnetic charge $Q_{\text{m}}=2q_{\text{m}}$ carried by a single elementary excitation in the dumbbell picture of dipolar and octupolar spin ices.
	\label{tab:magnetic_charge}}
\end{table}

The crossing term $J_{xz}$ is eliminated by a rotation
$\tau^z_i=S^z_i\cos\theta-S_i^x\sin\theta$ and $\tau^x_i=S_i^z\sin\theta+S_i^x\cos\theta$,
where the rotated spin-$1/2$ operator $\bm{S}_i$'s
give rise to an XYZ model with the extra dipole-dipole interaction,
\begin{equation}
\label{eq:XYZ}
	\begin{aligned}
		H&=\sum_{\braket{ij}}\left(\tilde{J}_xS_i^xS_j^x+\tilde{J}_yS_i^yS_j^y+\tilde{J}_zS_i^zS_j^z\right)\\
		&-\sum_i\left(\hat{\bm{z}}_i\cdot\hat{\bm{e}}\right)hg\mu_{\text{B}}\left(S_i^z\cos\theta-S_i^x\sin\theta\right)\\
		&+\frac{1}{2}\sum_{i,j}^{r_{ij}>r_{\text{nn}}}\frac{\mu_0 g^2\mu_{\text{B}}^2}{4\pi}\frac{\hat{\bm{z}}_i\cdot\hat{\bm{z}}_j-3(\hat{\bm{z}}_i\cdot\hat{\bm{r}}_{ij})(\hat{\bm{z}}_j\cdot\hat{\bm{r}}_{ij})}{r_{ij}^3}\\
		&\times\left(S_i^z\cos\theta-S_i^x\sin\theta\right)\left(S_j^z\cos\theta-S_j^x\sin\theta\right).
	\end{aligned}
\end{equation}
In the easy-axis limit of the XYZ model, the system is in the quantum spin ice regime and realizes a ground state as the $\mathrm{U}(1)$ spin liquid with a 3D $\mathrm{U}(1)$ topological order. Because of different symmetry properties of $S^x$, $S^y$, and $S^z$, types of $\mathrm{U}(1)$ topological orders are enriched \cite{huangQuantumSpinIces2014,liSymmetryEnrichedU12017}. When $\tilde{J}_x$ or $\tilde{J}_z$ dominates, the system realizes a dipolar $\mathrm{U}(1)$ spin liquid, while if $\tilde{J}^y$ dominates, it becomes an octupolar $\mathrm{U}(1)$ spin liquid. Below, we argue that the dipole-dipole interaction in the above equations provides a smoking-gun distinction in the magnetic charge of the magnetic monopole to distinguish the dipolar spin ice from the octupolar spin ice, and thereby distinguish the dipolar spin liquid from the octupolar spin liquid at the low-temperature limit. The main observation is summarized in Table~\ref{tab:magnetic_charge}.

It is straightforward to see that the long-range dipole-dipole interaction in Eq.~\eqref{eq:XYZ} only operates on the $S^x$ and $S^z$ components, not on the $S^y$ component. This observation immediately leads to the results in Table~\ref{tab:magnetic_charge}. To explain the physics and keep the generality, we assume $S^{\lambda}$ is the spin component along the easy axis, $J$ is the exchange interaction between the nearest neighbors $S^{\lambda}$ components and favors the degenerate spin ice configurations for the $S^{\lambda}$ components, and $J_\perp$ is the nearest-neighbor exchange for the spin components that are normal to the $S^{\lambda}$ component. The quantum mechanical tunneling events between different spin ice configurations are generated by the high-order perturbation of the transverse exchange, and this energy scale is set by $J_{\text{ring}}$. For instance, ${J_{\text{ring}} \sim \mathcal{O}(J_{\perp}^3/J^2)}$, and can be of even higher orders with other transverse exchanges~\cite{PhysRevB.69.064404}. It is the $J_{\text{ring}} $ interaction-driven quantum fluctuation that is responsible for the emergence of $\mathrm{U}(1)$ spin liquid at very low temperature with various emergent physical properties. In the temperature regime $T \gtrsim J_{\text{ring}}$, the quantum coherence is destroyed by the thermal fluctuations. Although the nongeneric collapse of quantum entanglement could occur~\cite{PhysRevB.87.205130}, the generic situation is the thermal crossover to the classical spin ice regime. In the classical spin ice regime, one can neglect the transverse exchange and keep only the nearest-neighbor Ising interaction and the dipole-dipole interaction between the $S^{\lambda}$ components. The resulting model is given as
\begin{equation}
\label{eq3}
	\begin{aligned}
		&H_{\text{CSI}}=\sum_{\braket{ij}}J S_i^{\lambda} S_j^{\lambda}
		-\sum_{i}(\hat{\bm{z}}_i\cdot\hat{\bm{e}})h\tilde{g}\mu_{\text{B}}S_i^{\lambda} \\
		&+\frac{1}{2}\sum_{i,j}^{r_{ij}>r_{\text{nn}}}\frac{\mu_0
		\tilde{g}^2\mu_{\text{B}}^2}{4\pi}\frac{\hat{\bm{z}}_i\cdot\hat{\bm{z}}_j
		-3(\hat{\bm{z}}_i\cdot\hat{\bm{r}}_{ij})(\hat{\bm{z}}_j\cdot\hat{\bm{r}}_{ij})}{r_{ij}^3}S_i^{\lambda}S_j^{\lambda}.
	\end{aligned}
\end{equation}
To unify the discussion of both dipolar case and octupolar case, we introduce an effective $g$ factor that depends on the easy axis $\lambda$ we choose (see Table~\ref{tab:magnetic_charge}). As Eq.~\eqref{eq1}, the first line of the above equation already includes the NN contribution from the dipole-dipole interaction.

\begin{figure}[htb]
	\includegraphics[width=0.9\linewidth]{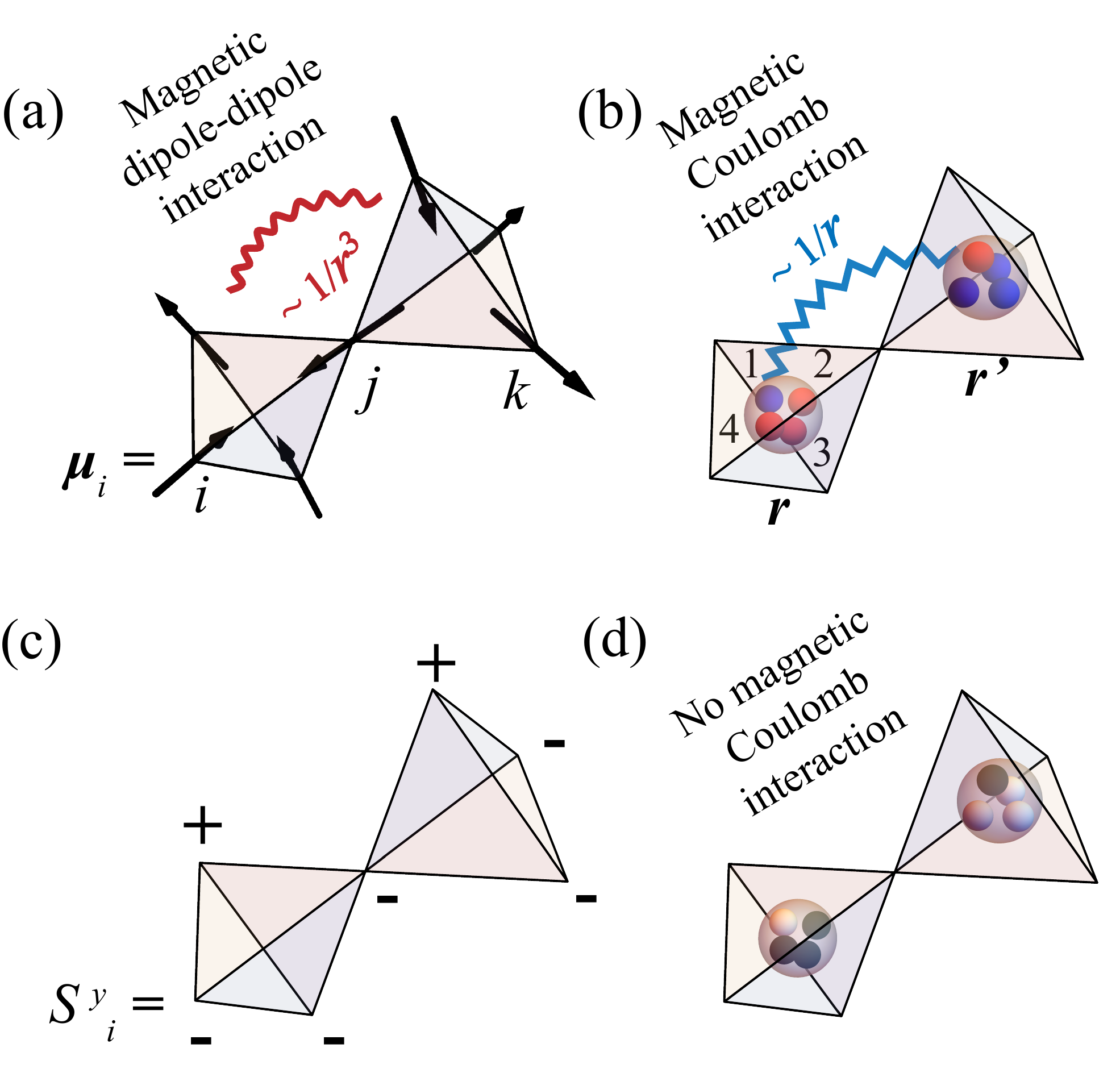}
    \caption{Illustration of the emergent monopole picture in classical spin ice. (a), (b) Dipolar spin ice. (a) A three-in-one-out tetrahedron (left) and a one-in-three-out tetrahedron (right) in the spin configuration. Black arrows represent the local dipole moments $\bm{\mu}_i$. They interact via the long-range $\sim r^{-3}$ dipole-dipole interaction.
    (b) In the corresponding dumbbell model, each spin is represented by a pair of oppositely charged magnetic monopoles $+ q_{\text{m}}$ (red balls) and $-q_{\text{m}}$ (blue balls) placed on the two adjacent tetrahedral centers. Four monopoles at the same tetrahedral center form a defect tetrahedron with magnetic charge $\pm Q_{\text{m}}$ (the big transparent balls).
    (c), (d) Octupolar spin ice. (c) The local moments are components of an octupole, represented by $+$ and $-$ signs for the eigenstates of $S^y_i$. No long-range dipole-dipole interaction exists.
    (d) In the analogous mapping, each spin is formally related to a positive monopole number (black balls) and a negative monopole number (white balls). A defect tetrahedron then carries a finite net monopole number but zero net magnetic charge.\label{fig:dumbbell}}
\end{figure}

\noindent\emph{Dumbbell picture}.---It was previously understood by Isakov, Moessner, and Sondhi that~\cite{isakovWhySpinIce2005}, the dipole-dipole interaction does not suppress the finite-temperature classical spin ice physics, though the ground state is finally driven to an antiferromagnetic order~\cite{PhysRevLett.87.067203}. An intuitive and neat dumbbell picture was then developed to incorporate the long-range dipole-dipole interaction into the classical spin ice of the nearest-neighbor Ising interaction~\cite{castelnovoMagneticMonopolesSpin2008}. The essential ingredient here is to view the Ising magnetic moment as the magnetic dipole of the effective or emergent magnetic monopole excitations of the classical spin ice, and the dipole-dipole interaction is well substituted by the Coulomb interaction between these magnetic monopoles. The effective magnetic charge ($q_{\text{m}}$) of the emergent magnetic monopole is then specified by the magnetic moment and the lattice constant of the underlying system~\cite{castelnovoMagneticMonopolesSpin2008}. Here, the temperature-dependent entropic interaction between the monopoles is neglected. In contrast, in the absence of the dipole-dipole interaction, the classical spin ice does not have these finite monopole charges from the dipole-dipole interaction. The ground state of the system simply demands the ice rule, i.e., every tetrahedron assumes a two-plus-two-minus configuration for the $S^{\lambda}$-components \cite{andersonOrderingAntiferromagnetismFerrites1956}. Breaking this ice rule gives rise to pointlike defects (known as magnetic monopoles) that reside at the tetrahedral centers. These monopoles are free to move without additional energy costs, as they experience no mutual interactions. Monopoles, in this case, have no effective magnetic charges.

To apply the dumbbell picture concretely to Eq.~\eqref{eq3}, we associate each Ising spin operator $S_i^{\lambda}$ with a pair of effective magnetic monopole operators $\eta_{\bm{r}}$ and $\eta_{\bm{r}'}$, located at the neighboring tetrahedral centers (see Fig.~\ref{fig:dumbbell}). The link $\bm{r}\bm{r}'$ is centered at the pyrochlore lattice site $i$ and the direction of ${\bm{r}'-\bm{r}}$ is along the unit vector $\hat{\bm{z}}_i$. The state $S^{\lambda}_i=+1/2$ ($S^{\lambda}_i=-1/2$) is mapped to the state ${\eta_{\bm{r}}=+1}$ and ${\eta_{\bm{r}'}=-1}$ (${\eta_{\bm{r}}=-1}$ and ${\eta_{\bm{r}'}=+1}$). Namely, $4S_i=\eta_{\bm{r}}-\eta_{\bm{r}'}$. The separation of these two monopoles is ${d=\sqrt{3/2}r_{\text{nn}}}$. To create a magnetic dipole moment $\tilde{\bm{\mu}}_i=\tilde{g}\mu_{\text{B}}S_i\hat{\bm{z}}_i$, each magnetic monopole should carry a magnetic charge $\eta_{\bm{r}}q_{\text{m}}$ with $q_{\text{m}}=\tilde{g}\mu_{\text{B}}/(2d)$. The dipole-dipole interaction can then be regarded as an approximation of the magnetic Coulomb interaction between these magnetic charges. As a result, Eq.~\eqref{eq3} can be replaced by an interacting monopole-gas model as
\begin{equation}
\label{eq4}
	H=\frac{1}{2}\sum_{\bm{r},\bm{r}'}\sum_{a,b=1}^4v_{\bm{r}\bm{r}',ab}
	-\frac{d}{2}\sum_{\bm{r}}\sum_{a=1}^{4}(\hat{\bm{z}}_a\cdot\hat{\bm{e}})h q_{\text{m}} \zeta_{\bm{r}}\eta_{\bm{r},a},
\end{equation}
with
\begin{equation}
	v_{\bm{r}\bm{r}',ab}=
	\begin{dcases}
		& \frac{\mu_0 q_{\text{m}}^2}{4\pi}\frac{\eta_{\bm{r},a}
		\eta_{\bm{r}',b}}{|\bm{r}-\bm{r}'|}, \quad \bm{r}\ne \bm{r}',\\
		& v_0 \eta_{\bm{r},a} \eta_{\bm{r},b}, \quad \bm{r}=\bm{r}',
	\end{dcases}
\end{equation}
where $\eta_{\bm{r},a}q_{\text{m}}$ is the $a$th magnetic charge at the tetrahedral center $\bm{r}$ and $v_0$ denotes a self-energy that satisfies
\begin{equation}
	v_0=\frac{J}{4}+\frac{1}{12}\left(4\sqrt{\frac{2}{3}}-1\right)\frac{\mu_0\tilde{g}^2\mu_{\text{B}}^2}{4\pi r_{\text{nn}}^3}.
\end{equation}
The self-energy is necessary to reproduce the nearest-neighbor exchange interaction $J$ (see the Supplemental Material \cite{SM} for details). Since tetrahedral centers form a diamond lattice with two sublattices, we have introduced a symbol $\zeta_{\bm{r}}=+1$ ($\zeta_{\bm{r}}=-1$) if $\bm{r}$ is the tetrahedral center in which $\hat{\bm{z}}_a$ is pointing outwards (inwards) from the center.

Since there are four magnetic monopoles at each tetrahedral center, one can regard them as a large monopole with a large magnetic charge. Defining a total monopole number operator
${N_{\bm{r}}=\sum_{a=1}^4 \eta_{\bm{r},a}}$, one rewrites the Hamiltonian in Eq.~\eqref{eq4} as
\begin{equation}\label{eq7}
	\begin{aligned}
		H&=\sum_{\bm{r}}\frac{1}{2}v_0 N_{\bm{r}}^2+\frac{1}{2}\sum_{\bm{r}\ne\bm{r}'}\frac{\mu_0 q_{\text{m}}^2}{4\pi}\frac{N_{\bm{r}}N_{\bm{r}'}}{|\bm{r}-\bm{r}'|}\\
		&-\frac{d}{2}\sum_{\bm{r}}\sum_{a=1}^{4}(\hat{\bm{z}}_a\cdot\hat{\bm{e}})h q_{\text{m}} \zeta_{\bm{r}}\eta_{\bm{r},a}.
	\end{aligned}
\end{equation}
The first term in Eq.~\eqref{eq7} is the self-energy carried by the magnetic monopole. The second term is the magnetic Coulomb interaction between these large monopoles. The third term is the magnetic potential created by the external magnetic field.

\noindent\emph{Distinguishing dipolar and octupolar cases.}---Based on the monopole gas model in \cref{eq7}, we proceed to distinguish the dipolar and octupolar spin ices. For the octupolar case, the relevant Ising moment in the model is the $S^y$ component, and there is no dipole-dipole interaction for this component. Thus, the effective magnetic charge of the magnetic monopole is zero with $q_{\text{m}}=0$, and \cref{eq7} is simply reduced to a classical spin ice
\begin{equation}
	H_{\text{oct}}=\sum_{\bm{r}}\frac{1}{2}v_0 N_{\bm{r}}^2,
\end{equation}
whose ground states satisfy ${N_{\bm{r}}=0}$ for all $\bm{r}$, demanded by the ice rule. This line of reasoning and approximation captures the monopole charge in the thermal spin ice regime, but fails to capture the effect of external magnetic fields in the octupolar spin ice. For the field effect, one could further extend the previous studies at zero-temperature to the finite-temperature regime~\cite{liSymmetryEnrichedU12017,PhysRevResearch.2.013334,PhysRevResearch.5.033169}.

For the dipolar spin ice, the situation becomes very different since the magnetic monopoles now carry the magnetic charges $q_{\text{m}}\ne 0$. Flipping a spin or an open string of spins creates two well-separated elementary excitations in the two defect tetrahedra. Each elementary excitation carries a magnetic charge ${Q_{\text{m}}=2q_{\text{m}}}$, and \cref{eq7} becomes a magnetic monopole gas with the magnetic Coulomb interaction. For $\text{Ce}_2\text{Sn}_2\text{O}_7$, the NN distance ${r_{\text{nn}}=2.66\mbox{\normalfont\AA}}$, and the magnetic dipole moment ${\mu\approx 1.18\mu_{\text{B}}}$ of $\text{Ce}^{3+}$ between $1\,\mathrm{K}$ and $10\,\mathrm{K}$ \cite{sibilleCandidateQuantumSpin2015}. If the Ising component in Eq.~\eqref{eq3} is the dipolar component $S^z$, one can estimate the magnetic charge $Q_{\text{m}}\approx 2.04\times 10^{-5}q_{\text{D}}$ by setting ${\theta = 0}$, where ${q_{\text{D}}=h/(\mu_0 e)}$ is the Dirac magnetic charge quantum. As we list in Table~\ref{tab:magnetic_charge}, the actual magnetic charge depends on the interaction and the $\theta$ angle. The direct measurement of the magnetic charge could actually tell us the value of $\theta$. If $\text{Ce}_2\text{Sn}_2\text{O}_7$ is located in the octupolar spin ice regime, we immediately have ${Q_{\text{m}}=0}$. Using the existing magnetic moments and lattice constants of other Ce pyrochlores, we proceed to evaluate the magnetic charges of the magnetic monopoles by assuming the system is in the dipolar spin ice regime with $S^z$ the Ising component and ${\theta =0}$. These results are summarized in Tab.~\ref{tab:materials}. Particularly, previously there was some neutron scattering evidence for the dipolar spin ice proposal for Ce$_2$Zr$_2$O$_7$~\cite{smithCaseQuantum2022,gao2024emergentphotonsfractionalizedexcitations}, and our idea can be used to determine the $\theta$ value in this system.

\begin{table}[htb]
	\begin{tabular}{ccccc}
		\hline\hline
		Material & $r_{\text{nn}}/\mbox{\normalfont\AA}$ & $\mu/\mu_{\text{B}}$ & $Q_{\text{m}}/10^{-5}q_{\text{D}}$ & Reference \\
		\hline
		$\text{Ce}_2\text{Sn}_2\text{O}_7$ & $2.66$ & $1.18$ & $2.04$ & \cite{sibilleCandidateQuantumSpin2015}\\
		$\text{Ce}_2\text{Zr}_2\text{O}_7$ & $2.55$ & $1.29$ & $2.34$ & \cite{smithCaseQuantum2022}\\
		$\text{Ce}_2\text{Hf}_2\text{O}_7$ & $2.68$ & $1.18$ & $2.03$ & \cite{poreeCrystalfieldStatesDefect2022a}\\
		$\text{Cd}\text{Er}_2\text{Se}_4$ & $4.05$ & $8.14$ & $9.24$ & \cite{gaoDipolarSpinIce2018}\\
		$\text{Cd}\text{Er}_2\text{S}_4$ & $3.96$ & $8.29$ & $9.64$ & \cite{gaoDipolarSpinIce2018}\\
		$\text{Mg}\text{Er}_2\text{Se}_4$ & $2.86$ & $8.30$ & $13.4$ & \cite{reig-i-plessisDeviationDipoleiceModel2019}\\
		\hline\hline
	\end{tabular}
	\caption{The nearest-neighbor distance $r_{\text{nn}}$, the typical magnetic moment $\mu$, and the estimated maximum magnetic charge $Q_{\text{m}}$ (assuming dipolar spin ice along $S^z$ with $\theta=0$) that can be reached by typical dipole-octupole materials. \label{tab:materials}}
\end{table}

This magnetic monopole charge generation in the thermal spin ice regime to distinguish the dipolar and octupolar cases does not require the ground state in the zero-temperature limit to be a spin liquid~\cite{huangQuantumSpinIces2014}, weakly ordered~\cite{yahneDipolarSpinIce2024}, Coulomb ferromagnet~\cite{PhysRevLett.108.037202}, nor Coulomb antiferromagnet~\cite{PhysRevResearch.5.L032027}. Therefore, this idea can be well adapted to other pyrochlore magnets with the dipole-octupole doublets. In fact, the Dy$^{3+}$ ion in the well-known classical spin ice material Dy$_2$Ti$_2$O$_7$ has the ground state doublet as the dipole-octupole doublet~\cite{huangQuantumSpinIces2014}, and the system is well in the dipolar spin ice regime. The monopole charge was known and measured as $1.25\times10^{-4}q_{\text{D}}$~\citep{castelnovoMagneticMonopolesSpin2008,dusadMagneticMonopoleNoise2019}. For the spinel CdEr$_2$X$_4$ (X = Se, S) and MgEr$_2$Se$_4$~\cite{gaoDipolarSpinIce2018,reig-i-plessisDeviationDipoleiceModel2019}, the Er$^{3+}$ ion was known to have the dipole-octupole ground state doublet. Experimentally, all of them were proposed as the dipolar spin ice. Again, assuming ${\theta=0}$, one can obtain the magnetic charge for these systems. Reference~\onlinecite{gaoDipolarSpinIce2018} has already obtained these values for CdEr$_2$X$_4$ (X = Se, S).

For the Nd-based pyrochlores, the physics is a bit complex. Taking Nd$_2$Zr$_2$O$_7$ for example, the Nd$^{3+}$ $\tau^z$ moment in Eq.~\eqref{eq1} experiences a magnetic moment fragmentation that has been well understood~\cite{petitObservationMagneticFragmentation2016,xuOrderOutCoulomb2020,xuAnisotropicExchangeHamiltonian2019,PhysRevB.94.104430}. This physics can be explained via Eq.~\eqref{eq:XYZ}. Although the $S^x$ interaction in Eq.~\eqref{eq:XYZ} has a strong antiferromagnetic interaction that gives rise to the spin ice physics, the weak ferromagnetic $S^z$ interaction actually gains more energy and produces an all-in-all-out magnetic order at very low temperature. Nevertheless, the spin-ice correlation from the $S^x$ interaction still controls the spin correlation properties that are measured by neutron scatterings. If we now apply the picture and reasoning of Eq.~\eqref{eq3}, the Ising component is $S^x$. Using ${\theta = 0.98\,\mathrm{rad}}$~\cite{xuAnisotropicExchangeHamiltonian2019} and ${\mu=1.26\mu_{\text{B}}}$, we find ${Q_{\text{m}} = 1.79\times 10^{-5}q_{\text{D}}}$. The magnetic charges of other Nd compounds (Nd$_2$Sn$_2$O$_7$ and Nd$_2$Hf$_2$O$_7$) with similar physics could be evaluated in the same fashion~\cite{PhysRevB.92.144423,PhysRevB.92.184418,PhysRevResearch.5.L032027}.

\noindent{\emph{Discussion.}}---Previously, we have argued that, the presence of the low-temperature thermal Hall effect of electric monopoles is an important property of dipolar $\mathrm{U}(1)$ spin liquid that differs fundamentally from the octupolar one that lacks this transport property~\cite{PhysRevResearch.2.013066}. Here, we focus on the magnetic monopole charge. The fundamental difference between the octupolar spin ice and the dipolar spin ice is whether the classical magnetic monopole excitations, connected to the spinons in the quantum spin liquid regime, carry a finite magnetic charge from the long-range dipole-dipole interaction. Therefore, one straightforward way to distinguish them is to directly measure the magnetic charge of monopole excitations. As magnetic charges can generate magnetic fields with a net divergence, a smoking-gun signature of magnetic monopoles is the quantized flux jump when they go through a superconducting ring \cite{cabreraFirstResultsSuperconductive1982}. Based on this phenomenon, it is possible to detect magnetic monopoles using a superconducting quantum interference device \cite{cabreraFirstResultsSuperconductive1982}. In the actual setting, because monopoles with positive and negative charges are generated simultaneously in the materials and their density fluctuates due to thermalization, the flux jump signature is stochastic \cite{klyuevMemoryEffectGenerationrecombination2019}. Thus, instead of measuring a single flux jump, one measures the flux noise \cite{dusadMagneticMonopoleNoise2019}. This technique has successfully measured the magnetic monopole noise \cite{dusadMagneticMonopoleNoise2019,klyuevMemoryEffectGenerationrecombination2019} in the classical spin ice materials such as $\text{Dy}_2\text{Ti}_2\text{O}_7$ and $\text{Ho}_2\text{Ti}_2\text{O}_7$~\cite{castelnovoMagneticMonopolesSpin2008}.

In addition, one can use regular magnetic and thermodynamic measurements to observe the indirect effect. In the monopole model, there has been a prediction of a first-order monopole liquid-gas transition for the dipolar spin ice in the $[111]$ magnetic field if the temperature is lower than a critical temperature~\cite{castelnovoMagneticMonopolesSpin2008}. This should work if the system is classical enough and has actually been observed in Dy$_2$Ti$_2$O$_7$, but may not apply well to the Ce pyrochlores, where other quantum mechanical terms start to play a more important role at low temperatures. We expect the Er compounds and Nd compounds to be promising systems to
realize this piece of physics. For the octupolar spin ice, the field effect is rich and complex, as the field itself creates the quantum mechanical process~\cite{liSymmetryEnrichedU12017}, and the monopole picture is expected to fail when the field becomes large.

To summarize, we have shown how to distinguish different symmetry enriched topological orders in dipole-octupole pyrochlores in the classical regime. Once the quantum coherence is destroyed by thermal fluctuations, the dipole-dipole interaction between dipolar spin components renders the elementary monopole excitation of the spin ice manifold a nonzero magnetic charge. Instead, if the system is in the octupolar spin ice regime, the elementary monopole excitations have a zero magnetic charge. The magnetic charge is the fundamental difference between dipolar and octupolar spin ices. The typical scale of these charges in dipole-octupole systems is ${Q_{\text{m}}\sim 10^{-5}q_{\text{D}}}$, and can be measured in the monopole noise experiments. More broadly, if certain properties of the quasiparticles of the symmetry enriched topological orders can persist in the classical regime, one could use classical physics to understand them. Furthermore, this strategy of diagnosing topological orders through their classical manifestations could potentially be extended to higher-rank $\mathrm{U}(1)$ spin liquids in breathing pyrochlore antiferromagnets~\cite{gresistaQuantumCoulombLiquids2026,yanRank2U1Spin2020}.

\noindent{\emph{Acknowledgment.}}---We acknowledge S\'{e}amus Davis for useful discussions. This work is supported by the National Natural Science Foundation of China (Grants No.~92565110 and No.~12574061) and by the Ministry of Science and Technology of China (Grant No.~2021YFA1400300).

\clearpage
\onecolumngrid

\setcounter{equation}{0}
\setcounter{section}{0}
\setcounter{figure}{0}
\setcounter{table}{0}
\renewcommand{\theequation}{S\arabic{equation}}
\renewcommand{\thesection}{ \Roman{section}}
\renewcommand{\theHequation}{S\arabic{equation}}
\renewcommand{\theHsection}{S\Roman{section}}
\renewcommand{\theHsubsection}{S\Roman{section}.\arabic{subsection}}
\renewcommand{\thefigure}{S\arabic{figure}}
\renewcommand{\theHfigure}{S\arabic{figure}}
\renewcommand{\thetable}{\arabic{table}}
\renewcommand{\theHtable}{S\arabic{table}}
\renewcommand{\tablename}{Supplementary Table}
\renewcommand{\bibnumfmt}[1]{[S#1]}
\renewcommand{\citenumfont}[1]{#1}

\begin{center}
\textbf{\large Supplemental material for}\\[0.75em]
\textbf{\large classical symmetry enriched topological orders and distinct monopole charges for dipole-octupole spin ices}\\[1em]
Pengwei Zhao\\
International Center for Quantum Materials, Peking University, Beijing 100871, China\\[0.75em]
Gang v.~Chen\\
International Center for Quantum Materials, Peking University, Beijing 100871, China\\
Tsung-Dao Lee Institute, Shanghai Jiao Tong University, Shanghai 201210, China\\
Collaborative Innovation Center of Quantum Matter, Beijing 100871, China\\
The University of Hong Kong, Shenzhen Institute of Research and Innovation, Shenzhen 518057, China\\
Beijing Key Laboratory of Quantum Devices, Peking University, Beijing 100871, China\\[1em]
\end{center}

\section{Pyrochlore lattice geometry and local frame}
\begin{figure}[htb]
        \centering
        \includegraphics[width=0.5\linewidth]{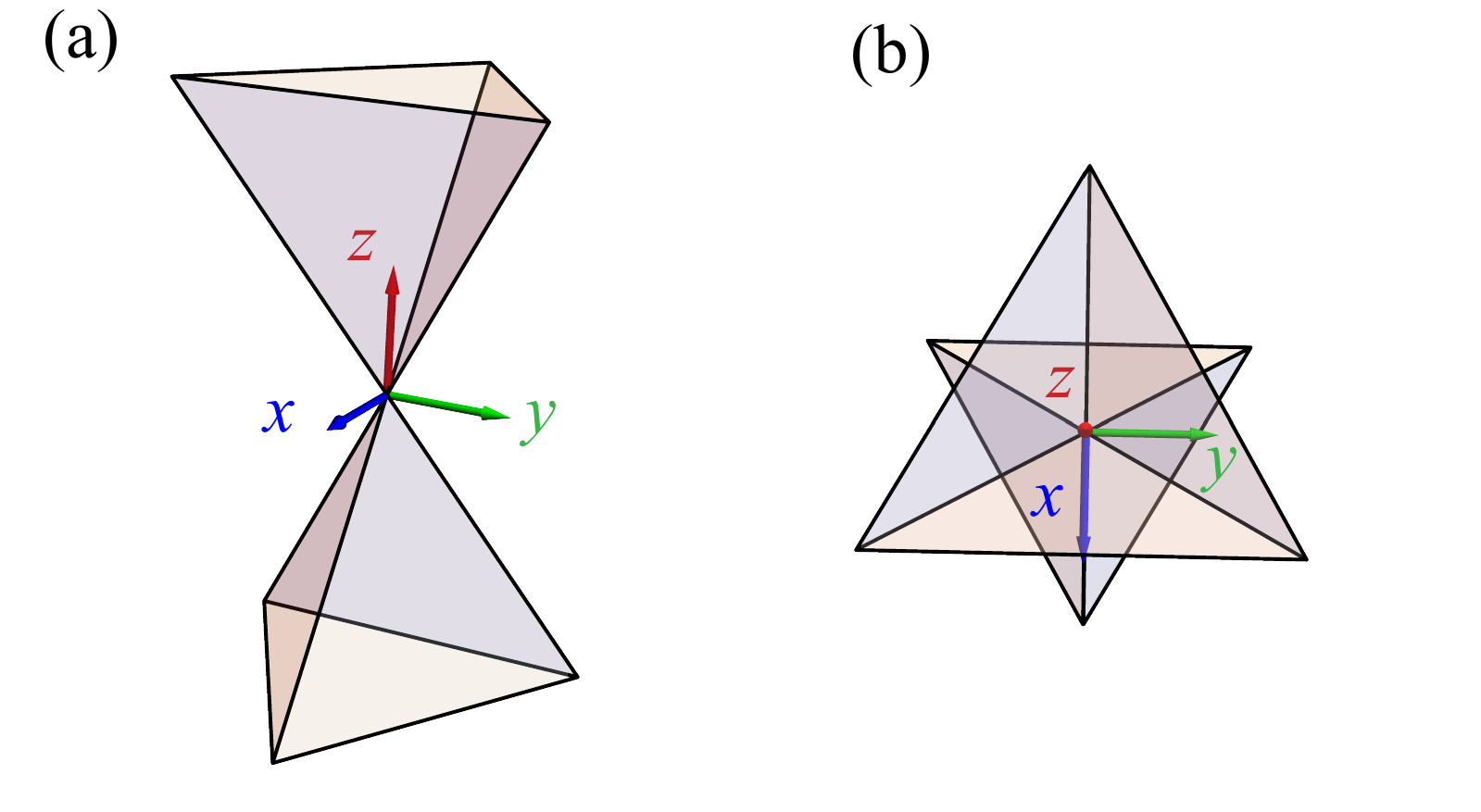}
        \caption{(a) Local coordinate frame defined at pyrochlore lattice sites. (b) Top view of (a).}
        \label{fig:frame}
    \end{figure}

    A pyrochlore lattice consists of corner-sharing tetrahedra, with $r_{\text{nn}}$ denoting the nearest-neighbor distance between sites. The lattice contains two types of tetrahedra: up and down tetrahedra. The up tetrahedra form a face-centered cubic (FCC) lattice with primitive vectors
    \begin{equation}
        \bm{a}_1=r_{\text{nn}}\left(0,1,1\right),\quad \bm{a}_2=r_{\text{nn}}\left(1,0,1\right),\quad \bm{a}_3=r_{\text{nn}}\left(1,1,0\right).
    \end{equation}
    Each up tetrahedron serves as a unit cell containing four sites labeled by $a=1,2,3,4$. Their position vectors relative to the tetrahedral center are
    \begin{equation}
        \bm{e}_1=\frac{r_{\text{nn}}}{2}\sqrt{\frac{3}{2}}(1,1,1),\quad \bm{e}_2=\frac{r_{\text{nn}}}{2}\sqrt{\frac{3}{2}}(1,-1,-1),\quad \bm{e}_3=\frac{r_{\text{nn}}}{2}\sqrt{\frac{3}{2}}(-1,1,-1),\quad \bm{e}_4=\frac{r_{\text{nn}}}{2}\sqrt{\frac{3}{2}}(-1,-1,1).
    \end{equation}
    At each site, a local coordinate frame is defined with the $z$-axis along $\bm{e}_a$:
    \begin{equation}
        \hat{\bm{z}}_1=\frac{1}{\sqrt{3}}(1,1,1),\quad \hat{\bm{z}}_2=\frac{1}{\sqrt{3}}(1,-1,-1),\quad \hat{\bm{z}}_3=\frac{1}{\sqrt{3}}(-1,1,-1),\quad \hat{\bm{z}}_4=\frac{1}{\sqrt{3}}(-1,-1,1).
    \end{equation}
    The local $y$-axes are defined as
    \begin{equation}
        \hat{\bm{y}}_1=\frac{1}{\sqrt{2}}(0,1,-1),\quad \hat{\bm{y}}_2=\frac{1}{\sqrt{2}}(-1,0,-1),\quad \hat{\bm{y}}_3=\frac{1}{\sqrt{2}}(-1,-1,0),\quad \hat{\bm{y}}_4=\frac{1}{\sqrt{2}}(-1,1,0).
    \end{equation}
    The local $x$-axes are then given by $\hat{\bm{x}}_a=\hat{\bm{y}}_a\times\hat{\bm{z}}_a$ for $a=1,2,3,4$. \Cref{fig:frame} illustrates the local frame for one site of an up tetrahedron.

	\section{A brief review of dipole-octupole doublets}
    Recent years have witnessed growing interest in pyrochlore quantum spin liquids, particularly in Ce-based pyrochlore magnets as promising platforms for realizing U(1) quantum spin liquids. In these materials, Ce$^{3+}$ ions form a pyrochlore lattice. Under spin-orbit coupling (SOC) and crystal electric field (CEF) effects, the Ce$^{3+}$ ions exhibit two-fold degenerate ground states known as dipole-octupole (DO) doublets. These DO doublets represent a special class of Kramers doublets whose degeneracy is protected by time-reversal symmetry. Specifically, the ground state wavefunctions are linear superpositions of total angular momentum states $J=5/2$ with $J_z=\pm3/2$, where the $z$-axis aligns with the local [111] direction. We define pseudospin operators as
    \begin{equation}
    \begin{aligned}
        \tau^x&=\frac{1}{2}\left(\ket{+3/2}\bra{-3/2}+\ket{-3/2}\bra{+3/2}\right), \\
        \tau^y&=\frac{1}{2}\left(-i\ket{+3/2}\bra{-3/2}+i\ket{-3/2}\bra{+3/2}\right), \\
        \tau^z&=\frac{1}{2}\left(\ket{+3/2}\bra{+3/2}-\ket{-3/2}\bra{-3/2}\right).
    \end{aligned}
    \end{equation}
    Under time reversal $\mathcal{T}=e^{-i \pi J_y}K$, where $K$ denotes complex conjugation, the doublets transform as
    \begin{equation}
        \mathcal{T}:\tau^x\to -\tau^x,\quad \tau^y\to -\tau^y,\quad \tau^z\to -\tau^z.
    \end{equation}
    The distinctive feature of DO doublets, setting them apart from conventional Kramers doublets, lies in the different multipolar characters of the pseudospin components. Under $D_{3d}$ point group operations and time reversal, the three components transform distinctly:
    \begin{itemize}
        \item $\tau^z$ transforms as a magnetic dipole moment along the local $z$-axis
        \item $\tau^x$ exhibits dipolar symmetry but physically corresponds to an octupolar magnetic moment
        \item $\tau^y$ transforms with pure octupolar symmetry
    \end{itemize}
    This classification stems from their transformation properties under $D_{3d}$ symmetry operations. The local $z$-axis serves as the 3-fold rotation axis. Under a 3-fold rotation $\mathcal{C}_3=e^{i\frac{2\pi}{3}J^z}$ about this axis, all pseudospin components remain invariant:
    \begin{equation}
        \mathcal{C}_3:\tau^x\to \tau^x,\quad \tau^y\to \tau^y,\quad \tau^z\to \tau^z.
    \end{equation}
    However, under mirror operations $\mathcal{M}$ and inversion $\mathcal{I}$, their transformations differ significantly. For a mirror plane $\mathcal{M}$ mapping $+x$ and $+y$ to $-y$ and $-x$ respectively:
    \begin{equation}
        \mathcal{M}: \tau^x \rightarrow -\tau^x, \quad \tau^y \rightarrow \tau^y, \quad \tau^z \rightarrow -\tau^z
    \end{equation}
    Under inversion $\mathcal{I}$, all components remain invariant:
    \begin{equation}
        \mathcal{I}: \tau^x \rightarrow \tau^x, \quad \tau^y \rightarrow \tau^y, \quad \tau^z \rightarrow \tau^z
    \end{equation}
    Thus, only $\tau^z$ transforms as a conventional magnetic dipole moment. The magnetic dipole moment $\bm{\mu}$ of DO doublets consequently arises solely from $\tau^z$. Explicitly, projecting the magnetic moment operator $\bm{\mu}\sim\mathcal{P}\bm{J}\mathcal{P}$ onto the DO doublet subspace yields
    \begin{equation}
        \bm{\mu}_i = \mu_B\left[g_{xy}(\hat{\bm{x}}_i\hat{\tau}_i^x + \hat{\bm{y}}_i\hat{\tau}_i^y) + g \hat{\bm{z}}_i\hat{\tau}_i^z\right]
    \end{equation}
    with $g\neq 0$ and $g_{xy}=0$ for DO doublets. This means only $\tau^z$ contributes to the conventional magnetic dipole moment detectable by techniques like neutron scattering and magnetic susceptibility.

    The $\tau^x$ and $\tau^y$ components correspond to magnetic octupole moments. While $\tau^x$ transforms with dipolar symmetry under $D_{3d}$, it physically represents an octupolar moment. The octupolar character of both components becomes evident upon projecting octupolar operators onto the DO doublet subspace:
    \begin{equation}
        \tau^x\propto \mathcal{P}\left[J_x^3-\frac{1}{3}\left(J_xJ_y^2+J_yJ_xJ_y+J_y^2J_x\right)\right]\mathcal{P},\quad \tau^y\propto \mathcal{P}\left[J_y^3-\frac{1}{3}\left(J_yJ_x^2+J_xJ_yJ_x+J_x^2J_y\right)\right]\mathcal{P}.
    \end{equation}
    In summary, the magnetic moments of DO doublets comprise both dipole ($\tau^z$) and octupole ($\tau^x$, $\tau^y$) components.

    \section{Technical details of dumbbell picture}
    \subsection{Model construction}
    In this section, we clarify the details for the derivations and dumbbell picture in the main text. The entire Hamiltonian of the pyrochlore spin ice with DO doublets contains three fundamental interactions: superexchange between local moments, external magnetic fields, and interactions between magnetic moments. In principle, there are two different but equivalent ways of writing the interaction between the local moments. The first way is to write the superexchange interaction plus the dipole-dipole interaction starting from the nearest neighbors. Because 4f electrons are pretty localized, one only needs to keep the nearest neighbor superexchange. The second way is to write all symmetry-allowed interactions between the local moments starting from the nearest neighbors. The nearest-neighbor symmetry-allowed interaction is the XYZ model, and this interaction includes the contributions from the superexchange and the dipole-dipole interaction. From the second neighbors, it is expected that the superexchange contribution becomes negligible due to the localized 4f electrons, and one can simply keep the dipole-dipole interaction as the major contribution. This second way of expressing the interaction is what we have adopted in the main text.

    Here, for self-consistency and as a complement, we follow the first approach. For the superexchange terms, since $\tau^x$ and $\tau^z$ share the same symmetry while $\tau^y$ behaves differently, one can write down the most generic form of superexchange $H_{\text{ex}}$ allowed by symmetries,
    \begin{equation}
        H_{\text{ex}}=\sum_{\braket{ij}}\left[J_x\tau^x_i\tau^x_j+J_y\tau^y_i\tau^y_j+J_z\tau^z_i\tau^z_j+J_{xz}\left(\tau^x_i\tau^z_j+\tau^z_i\tau^x_j\right)\right].
    \end{equation}
    For an external magnetic field $\bm{h}=h\hat{\bm{e}}$, where $\hat{\bm{e}}$ is a unit vector that specifies the direction of $\bm{h}$, the magnetic dipole moment directly couples with the external field $\bm{h}$, while the magnetic octupole moment couples with the second-order spatial derivative of $\bm{h}$. Therefore, for a uniform magnetic field $\bm{h}$, we only have
    \begin{equation}
        H_{\text{field}}=-\sum_{i}\left(\hat{\bm{z}}_i\cdot\hat{\bm{e}}\right)hg\mu_{\text{B}}\tau^z_i.
    \end{equation}
    As for interactions between magnetic moments, dipole moments couple via a $\sim 1/r^3$ dipolar interaction, while interactions between octupole moments are of much higher order and can be ignored. Therefore,
    \begin{equation}
        H_{\text{int}}=\frac{1}{2}\sum_{i\ne j}\frac{\mu_0g^2\mu_{\text{B}}^2}{4\pi}\frac{\hat{\bm{z}}_i\cdot\hat{\bm{z}}_j-3\left(\hat{\bm{z}}_i\cdot\hat{\bm{r}}_{ij}\right)\left(\hat{\bm{z}}_j\cdot\hat{\bm{r}}_{ij}\right)}{r_{ij}^3}\tau^z_i\tau^z_j.
    \end{equation}
    The total Hamiltonian is then given by $H=H_{\text{ex}}+H_{\text{field}}+H_{\text{int}}$.

    In principle, $H_{\text{int}}$ includes all dipolar interactions between all sites. However, in experimental measurements, dipolar interactions between nearest neighbors cannot be distinguished from the superexchange $J_z$. Therefore, it is convenient to absorb them into $J_z$. This is possible because dipolar interactions between nearest neighbors are isotropic. Namely, for a nearest neighbor $i$ and $j$, we have
    \begin{equation}
        H_{\text{int}}^{\text{nn}}=\sum_{\braket{ij}}\frac{\mu_0 g^2\mu_{\text{B}}^2}{4\pi r_{\text{nn}}^3}\left[\hat{\bm{z}}_i\cdot\hat{\bm{z}}_j+3\left(\frac{\hat{\bm{z}}_i\cdot\hat{\bm{z}}_j-1}{|\hat{\bm{z}}_j-\hat{\bm{z}}_i|}\right)^2\right]\tau^z_i\tau^z_j=\sum_{\braket{ij}}\frac{5}{3}\frac{\mu_0 g^2\mu_{\text{B}}^2}{4\pi r_{\text{nn}}^3}\tau^z_i\tau^z_j.
    \end{equation}
    Therefore, we redefined
    \begin{equation}
        J_z\to J_z-\frac{5}{3}\frac{\mu_0 g^2\mu_{\text{B}}^2}{4\pi r_{\text{nn}}^3},
    \end{equation}
    which removes the nearest-neighbor interactions from $H_{\text{int}}$,
    \begin{equation}
        H_{\text{int}}\to \frac{1}{2}\sum_{i\ne j}^{r_{ij}>r_{\text{nn}}}\frac{\mu_0g^2\mu_{\text{B}}^2}{4\pi}\frac{\hat{\bm{z}}_i\cdot\hat{\bm{z}}_j-3\left(\hat{\bm{z}}_i\cdot\hat{\bm{r}}_{ij}\right)\left(\hat{\bm{z}}_j\cdot\hat{\bm{r}}_{ij}\right)}{r_{ij}^3}\tau^z_i\tau^z_j.
    \end{equation}
    The Hamiltonian in the main text is then recovered.

    \subsection{Mapping dipolar interaction to Coulomb interaction}
\begin{figure}[htb]
        \centering
        \includegraphics[width=0.3\linewidth]{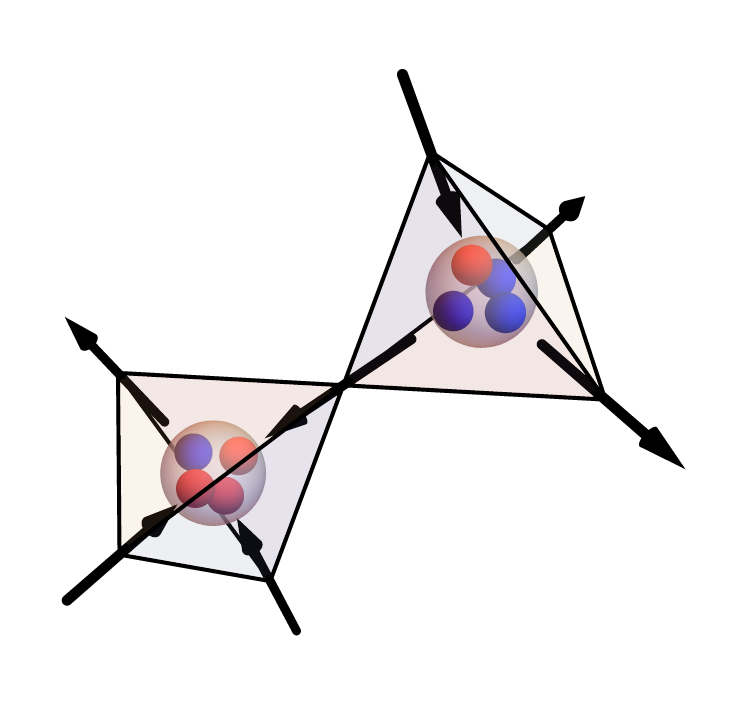}
        \caption{The key in the dumbbell picture is to replace local dipole moments by pairs of magnetic monopoles. Black arrows indicate the local magnetic moments. They can be approximated by magnetic charges located at the tetrahedron centers. In the figure, we show these monopoles at the same tetrahedron center by a shift to avoid overlap.}
        \label{fig:sm_dumbbell}
    \end{figure}

    By defining a rotated spin operator $\bm{S}$ and a rotation $\tau^z_i=S^z_i\cos\theta-S^x_i\sin\theta$ and $\tau^x_i=S^z_i\sin\theta+S^x_i\cos\theta$ with $\theta$ a solution of $J_{xz}(\cos^2\theta-\sin^2\theta)+(J_x-J_z)\sin\theta\cos\theta=0$, we can eliminate the $J_{xz}$ cross term. Under the classical limit, we obtain the classical spin ice model
    \begin{equation}\label{eq:csi}
        H_{\text{CSI}}=\sum_{\braket{ij}}JS^\lambda_i S^\lambda_j-\sum_i(\hat{\bm{z}}_i\cdot\hat{\bm{e}})h\tilde{g}\mu_{\text{B}} S^\lambda_i+\frac{1}{2}\sum_{i,j}^{r_{ij}>r_{\text{nn}}}\frac{\mu_0\tilde{g}^2\mu_{\text{B}}^2}{4\pi}\frac{\hat{\bm{z}}_i\cdot\hat{\bm{z}}_j-3\left(\hat{\bm{z}}_i\cdot\hat{\bm{r}}_{ij}\right)\left(\hat{\bm{z}}_j\cdot\hat{\bm{r}}_{ij}\right)}{r_{ij}^3}S^\lambda_iS^\lambda_j,
    \end{equation}
    where $\tilde{g}$ is an effective $g$ factor depending on the classical axis $\lambda$,
    \begin{equation}
        \tilde{g}=
        \begin{cases}
        & -g\sin\theta, \quad \lambda=x, \\
        & 0, \quad \lambda=y, \\
        & g\cos\theta, \quad \lambda=z.
        \end{cases}
    \end{equation}
    Eq.~\eqref{eq:csi} can be regarded as a model of magnetic dipole moments $\tilde{\bm{\mu}}_i=\hat{\bm{z}}_i\tilde{g}\mu_{\text{B}}S^\lambda_i$. To apply the dumbbell picture, we approximate each magnetic dipole moment $\tilde{\bm{\mu}}_i$ by two magnetic monopoles with opposite magnetic charges. For a magnetic monopole with magnetic charge $q_{\text{m}}=\tilde{g}\mu_{\text{B}}/(2d)$ located at a tetrahedron center, and another magnetic monopole with magnetic charge $-q_{\text{m}}$ located at the adjacent tetrahedron center, the magnetic dipole moment can be reproduced (see Fig.~\ref{fig:sm_dumbbell}). Here, $d=\sqrt{3/2}r_{\text{nn}}$ is the distance between two adjacent tetrahedron centers.

    For clarity, we ignore the external magnetic field at this stage. The Hamiltonian of such a monopole model is given by
    \begin{equation}\label{eq:monopoles}
        H=\frac{1}{2}\sum_{\bm{r},\bm{r}'}\sum_{a,b=1}^4v_{\bm{r}\bm{r}',ab},
    \end{equation}
    \begin{equation}
        v_{\bm{r}\bm{r}',ab}=
        \begin{dcases}
            & \frac{\mu_0 q_{\text{m}}^2}{4\pi}\frac{\eta_{\bm{r},a}\eta_{\bm{r}',b}}{|\bm{r}-\bm{r}'|},\quad \bm{r}\ne \bm{r}', \\
            & v_0\eta_{\bm{r},a}\eta_{\bm{r},b},\quad \bm{r}=\bm{r}',
        \end{dcases}
    \end{equation}
    where $\bm{r},\bm{r}'$ label the tetrahedron centers and $a,b=1,2,3,4$ indicates the four monopoles contained in each tetrahedron. $\eta_{\bm{r},a}$ denotes the sign of the $a$th magnetic charge at $\bm{r}$. The first line of $v_{\bm{r}\bm{r}',ab}$ is the magnetic Coulomb interaction between these magnetic monopoles in different tetrahedra, and the second line comes from the interaction between those in the same tetrahedron.

    \begin{figure}[htb]
        \centering
        \includegraphics[width=0.8\linewidth]{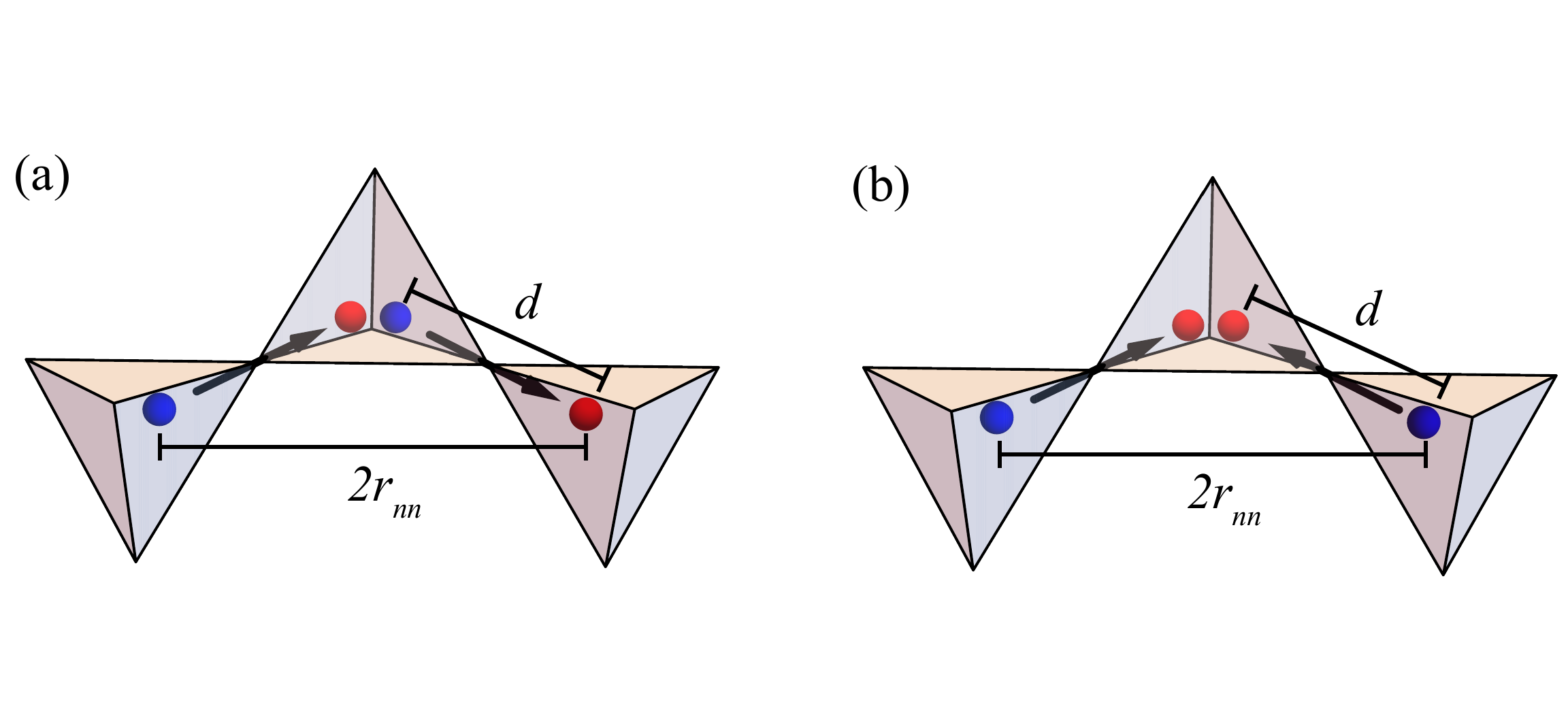}
        \caption{Two possible dipole configurations and the corresponding monopole configurations. Blue balls represent $\eta_{\bm{r},a}=-1$ and red balls represent $\eta_{\bm{r},a}=1$.}
        \label{fig:self_energy}
    \end{figure}

    The self-energy $v_0$ is determined such that Eq.~\eqref{eq:monopoles} can correctly reproduce Eq.~\eqref{eq:csi} for any configuration of dipole moments. Concretely, let's consider the interaction between two nearest neighboring dipole moments. Fig.~\ref{fig:self_energy} shows two possible dipole configurations and their corresponding monopole configurations. For Fig.~\ref{fig:self_energy}(a), the energy of dipoles is $-J/4$. It should equal the energy of monopoles, which gives
    \begin{equation}
        -\frac{J}{4}=-v_0-\frac{\mu_0q_{\text{m}}^2}{8\pi r_{\text{nn}}}.
    \end{equation}
    For the same reason, Fig.~\ref{fig:self_energy}(b) gives
    \begin{equation}
        \frac{J}{4}=v_0-\frac{\mu_0q_{\text{m}}^2}{\pi d}+\frac{\mu_0q_{\text{m}}^2}{8\pi r_{\text{nn}}}.
    \end{equation}
    A combination of these two equations gives
    \begin{equation}
        v_0=\frac{J}{4}+\frac{\mu_0q_{\text{m}}^2}{2\pi d}-\frac{\mu_0q_{\text{m}}^2}{8\pi r_{\text{nn}}}=\frac{J}{4}+\left(4\sqrt{\frac{2}{3}}-1\right)\frac{\mu_0 q_{\text{m}}^2}{8\pi r_{\text{nn}}}=\frac{J}{4}+\frac{1}{12}\left(4\sqrt{\frac{2}{3}}-1\right)\frac{\mu_0\tilde{g}^2\mu_{\text{B}}^2}{4\pi r_{\text{nn}}^3}.
    \end{equation}
    Therefore, up to the dipole order, Eq.~\eqref{eq:monopoles} is a good approximation of Eq.~\eqref{eq:csi}. However, there are corrections of interactions with orders higher than $1/r^{5}$, since replacing a pure dipole by two monopoles introduces higher-order moments. Those interactions decay very fast over the pyrochlore lattice, and hence, can be ignored.

\end{document}